\documentclass[onecolumn,showpacs]{revtex4}

\usepackage{graphicx}
\usepackage{dcolumn}
\usepackage{bm}

\input epsf

\begin{document}

\title{\Large Density Evolution in the New Modified Chaplygin Gas Model}

\author{\bf Surajit Chattopadhyay$^1$\footnote{surajit$_{_{-}}2008$@yahoo.co.in} and
Ujjal Debnath$^2$\footnote{ujjaldebnath@yahoo.com ,
ujjal@iucaa.ernet.in}}

\affiliation{$^1$Department of Information Technology, Pailan College of Management and Technology, Bengal Pailan Park, Kolkata-700 104, India.\\
$^2$Department of Mathematics, Bengal Engineering and Science
University, Shibpur, Howrah-711 103, India.}

\date{\today}

\begin{abstract}
In this paper, we have considered new modified Chaplygin gas
(NMCG) model which interpolates between radiation at early stage
and $\Lambda$CDM at late stage. This model is regarded as a
unification of dark energy and dark matter (with general form of
matter). We have derived the density parameters from the equation
of motion for the interaction between dark energy and dark
matter. Also we have studied the evolution of the various
components of density parameters.
\end{abstract}


\maketitle

\section{\normalsize\bf{Introduction}}
The standard cosmological model (SCM) can only describe
decelerated universe models and so cannot reproduce the results
coming from the recent type Ia supernovae observations upto about
$z\sim 1$ [1] which favour an accelerated current universe. But
the SCM can give a satisfactory explanation to other
observational properties of the present universe. The recent
extensive search for a matter field has given rise to the concept
of an accelerated expansion for the universe. This type of matter
is called Q-matter. This Q-matter can behave like a cosmological
constant [2] by combining +ve energy density and negative
pressure. So there must be this Q-matter either neglected or
unknown responsible for this accelerated universe. At  the
present epoch, a  lot of works has been done to  solve this
quintessence problem and most popular candidates for Q-matter
has  so  far been a scalar field having a  potential which
generates  a sufficient negative pressure. Furthermore,
observations reveal that the unknown form of matter properly
referred to as the `{\it dark energy}' accounts for almost $70\%$
of the universe. This is confirmed by the very recent WMAP data
[3]. A large number of possible candidates for this `{\it dark
energy}' component has already been proposed and their behaviour
have been studied extensively [4]. In order to explain the nature
of dark energy, many models have been proposed, such as,
k-essence [5, 6], tachyon [7], phantom [8], spintessence [9],
etc. So far, the theoretical probe of dark energy focuses mainly
on the evolution of the dark energy density or the equation of
state. The current astronomical observations
data cannot determine completely the nature of dark energy [10].\\

Another alternative candidate for Q-matter is exotic type of
fluid $-$ the so-called Chaplygin gas which obeys the equation of
state $p=-\tilde{A}/\rho, (\tilde{A}>0)$ [11], where $p$ and
$\rho$ are respectively the pressure and energy density.
Subsequently, the above equation was generalized to the form
$p=-\tilde{A}/\rho^{\alpha}, 0\le \alpha \le 1$ [12, 13] and
recently it was modified to the form
$p=\gamma\rho-\tilde{A}/\rho^{\alpha}, (\gamma>0)$ [14, 15], which
is known as {\it Modified Chaplygin Gas} (MCG). This model
represents the evolution of the universe starting from the
radiation era to the $\Lambda$CDM model. Recently Guo and Jhang
[16] proposed variable Chaplygin gas model where $\tilde{A}$ is a
positive function of the cosmological scale factor `$a$' i.e.,
$\tilde{A}=\tilde{A}(a)$. This assumption is reasonable since
$\tilde{A}(a)$ is related to the scalar potential if we take the
Chaplygin gas Born-Infeld scalar field [17]. Plethora of
literatures is available on the study of variable Chaplygin gas
model [18]. Also, New Generalized Chaplygin Gas
(NGCG) model have been discussed by Jhang et al [19].\\

Another type of dark energy roughly includes quiessence (or
$X$-matter) [20]. The quiessence or $X$-matter component is simply
characterized by a constant, non-positive equation of state
$w_{_{X}}$, where $w_{_{X}}$ is the ratio of pressure and density
of this $X$-matter. For accelerating universe, $w_{_{X}}<-1/3$.
For a normal scalar field with potential in FRW background with
the presence of cold dark matter (CDM), the bound of $w_{_{X}}$
would be $-1<w_{_{X}}<0$. Also $w_{_{X}}<-1$ is possible in the
framework of $X$CDM ($X$-matter with CDM) by fitting the SNe Ia
data. From observational data, the range of the equation of state
$w_{_{X}}$ of dark energy has been determined as
$-1.46<w_{_{X}}<-0.78$. Now MCG generalizes to accommodate any
possible $X$-type dark energy. So we consider here New Modified
Chaplygin Gas (NMCG) model as a scheme for unification of
$X$-type dark energy and dark matter. The new feature of this
model is that it behaves as a dark matter (radiation) at early
stage and $X$-type dark energy at late stage. We will show that
this model is a kind of interacting
$X$CDM system.\\

The organization of the paper is as follows: In section II, we
have considered New Modified Chaplygin Gas (NMCG) model and find
out the energy densities for dark energy and dark matter of this
model in FRW universe. Section III describes the interaction
between dark energy and dark matter and the evolution of density
parameters for dark energy and dark matter. Finally, the paper
ends with concluding remarks in section IV.\\

\section{\normalsize\bf{FRW model and NMCG model}}

The metric of a homogeneous and isotropic universe in the FRW
model is

\begin{equation} ds^{2}= dt^{2}-a^{2}(t)\left[\frac{dr^{2}}{1-kr^{2}}+
r^{2}(d\theta^{2}+\sin^{2}\theta d\phi^{2})\right]
\end{equation}

where $a(t)$ is the scale factor and $k$ (= 0, $\pm{1}$) is the
curvature scalar. The Einstein field equations are

\begin{equation}
\frac{\dot{a}^{2}}{a^{2}}+\frac{k}{a^{2}}=\frac{1}{3}~\rho
\end{equation}

\begin{equation}
\frac{\ddot{a}}{a}=-\frac{1}{6}~(\rho+3p)
\end{equation}

where $\rho$ and $p$ are energy density and isotropic pressure
respectively (choosing $8\pi G=c=1$).\\

The equation of state for NMCG model is

\begin{equation}
p_{_{ch}} = \gamma\rho_{_{ch}}-
\frac{\tilde{A}(a)}{\rho_{_{ch}}^{\alpha}}~~,~\gamma
> 0, ~0\leq \alpha \leq 1
\end{equation}

where $\tilde{A}(a)$ is a function that depends upon the scale
factor of the universe.\\

In the framework of FRW cosmology, considering the exotic
background fluid, the NMCG is described by the equation of state

\begin{equation}
\dot{\rho}_{_{ch}}+ 3\frac{\dot{a}}{a}(\rho_{_{ch}} + p_{_{ch}})=
0
\end{equation}

We know that the exotic background fluid smoothly interpolates
between a dark matter dominated phase $\rho\sim a^{-3(1+\gamma)}$
to dark energy dominated phase $\rho\sim a^{-3(1+w_{_{X}})}$ where
$w_{_{X}} (<-1/3)$ is the ratio of pressure and energy density of
$X$-matter (dark energy). For this purpose, without any loss of
generality, we consider the function $\tilde{A}(a)$ has in the
form [19]

\begin{equation}
\tilde{A}(a)=-w_{_{X}}A~a^{-3(1+w_{_{X}})(1+\alpha)}~,~~A>0
\end{equation}

so that using equations (4)-(6), the energy density of the NMCG
can be expressed as

\begin{equation}
\rho_{_{ch}}=\left[\frac{w_{_{X}}}{w_{_{X}}-\gamma}~A~a^{-3(1+w_{_{X}})
(1+\alpha)}+B~a^{-3(1+\gamma)(1+\alpha)}
\right]^{\frac{1}{1+\alpha}}
\end{equation}

Now NMCG scenario involves an interacting $X$CDM system. For
showing this, we first decompose the NMCG fluid into two
components i.e., dark energy and dark matter components (i.e.,
$\rho_{_{ch}}=\rho_{_{X}}+\rho_{_{dm}}$). There are several works
on such decomposition procedure in Chaplygin gas model [19, 21].
So we can obtained the densities of dark energy and dark matter
components respectively as

\begin{equation} \rho_{_{X}}=\frac{\frac{w_{_{X}}}{w_{_{X}}-\gamma}~A~a^{-3(1+w_{_{X}})
(1+\alpha)}+\frac{\gamma}{w_{_{X}}}~B~a^{-3(1+\gamma)(1+\alpha)}}{\left[\frac{w_{_{X}}}{w_{_{X}}-\gamma}~A~a^{-3(1+w_{_{X}})
(1+\alpha)}+B~a^{-3(1+\gamma)(1+\alpha)}
\right]^{\frac{1}{1+\alpha}}}
\end{equation}
and
\begin{equation}
\rho_{_{dm}}=\frac{\frac{w_{_{X}}-\gamma}{w_{_{X}}}~B~a^{-3(1+\gamma)(1+\alpha)}}{\left[\frac{w_{_{X}}}{w_{_{X}}-\gamma}~A~a^{-3(1+w_{_{X}})
(1+\alpha)}+B~a^{-3(1+\gamma)(1+\alpha)}
\right]^{\frac{1}{1+\alpha}}}
\end{equation}

From these two expressions one obtains the scaling behaviour of
the energy densities

\begin{equation}
\frac{\rho_{_{dm}}}{\rho_{_{X}}}=\frac{\frac{w_{_{X}}-\gamma}{w_{_{X}}}~B~a^{-3(1+\gamma)(1+\alpha)}}{\frac{w_{_{X}}}{w_{_{X}}-\gamma}~A~a^{-3(1+w_{_{X}})
(1+\alpha)}+\frac{\gamma}{w_{_{X}}}~B~a^{-3(1+\gamma)(1+\alpha)}}
\end{equation}

Parameters $A$ and $B$ can be expressed using current cosmological
observations. It is easy to get

\begin{equation}
A+B=\rho_{_{ch0}}^{\eta}
\end{equation}

where, $\eta=1+\alpha$ is used to characterize the interaction for
simplicity, thus we have

\begin{equation}
A=A_{s}\rho_{_{ch0}}^{\eta}~~,~~~~B=(1-A_{s})\rho_{_{ch0}}^{\eta}
\end{equation}

where $A_{s}$ is a dimensional parameter. Using equations (10)
and (12), one gets

\begin{equation}
A_{s}=\frac{\rho_{_{X0}}(\gamma-w_{_{X}})^{2}+\rho_{_{dm0}}\gamma(\gamma-w_{_{X}})
}{(w_{_{X}}^{2}-\gamma
w_{_{X}}+\gamma^{2})\rho_{_{dm0}}+(\gamma-w_{_{X}})^{2}\rho_{_{X0}}
}
\end{equation}

Here, we have assumed that the universe is flat. Hence the NMCG
energy density can be expressed as

\begin{equation}
\rho_{_{ch}}=\rho_{_{ch0}}a^{-3(1+\gamma)}\left[1-A_{s}
\left(1-\frac{w_{_{X}}}{w_{_{X}}-\gamma}~a^{-3\eta(w_{_{X}}-\gamma)}
\right) \right]^{\frac{1}{\eta}}
\end{equation}

Making use of (7), (8), (9) and (14), the energy densities of
dark energy and dark matter can be re-expressed as

\begin{equation}
\rho_{_{X}}=\rho_{_{X0}}a^{-3(1+\gamma)}
\frac{\frac{\gamma}{w_{_{X}}}~(1-A_{s})+
\frac{w_{_{X}}A_{s}}{w_{_{X}}-\gamma}~a^{-3\eta(w_{_{X}}-\gamma)}}{\frac{\gamma}{w_{_{X}}}~(1-A_{s})+
\frac{w_{_{X}}A_{s}}{w_{_{X}}-\gamma} }
 \left(1+\frac{\gamma A_{s}}{w_{_{X}}-\gamma}
\right)^{1-\frac{1}{\eta}}\left[1-A_{s}
\left(1-\frac{w_{_{X}}}{w_{_{X}}-\gamma}~a^{-3\eta(w_{_{X}}-\gamma)}
\right) \right]^{\frac{1}{\eta}-1}
\end{equation}

\begin{equation}
\rho_{_{dm}}=\rho_{_{dm0}}a^{-3(1+\gamma)}\left(1+\frac{\gamma
A_{s}}{w_{_{X}}-\gamma} \right)^{1-\frac{1}{\eta}}\left[1-A_{s}
\left(1-\frac{w_{_{X}}}{w_{_{X}}-\gamma}~a^{-3\eta(w_{_{X}}-\gamma)}
\right) \right]^{\frac{1}{\eta}-1}
\end{equation}

\section{\normalsize\bf{Interaction Between Dark Matter and Dark Energy}}

The whole NMCG fluid satisfies the energy conservation, but dark
energy and dark matter components do not obey the energy
conservation separately; they interact with each other. We
portray the interaction through an energy exchange term $Q$. The
equations of motion for dark energy and dark matter can be
written as

\begin{equation}
\dot{\rho}_{_{X}}+3H(1+w_{_{x}})\rho_{_{X}}=Q
\end{equation}

\begin{equation}
\dot{\rho}_{_{dm}}+3H(1+\gamma)\rho_{_{dm}}=-Q
\end{equation}

where, $H=\frac{\dot{a}}{a}$ represents the Hubble parameter. We
define the effective equations of state for dark energy and dark
matter through the parameters

\begin{equation}
w_{_{X}}^{(e)}=w_{_{X}}-\frac{Q}{3H\rho_{_{X}}}
\end{equation}

\begin{equation}
w_{_{dm}}^{(e)}=\gamma+\frac{Q}{3H\rho_{_{dm}}}
\end{equation}

The equations of dark energy and dark matter can be re-expressed
as

\begin{equation}
\dot{\rho}_{_{X}}+3H(1+w_{_{X}}^{(e)})\rho_{_{X}}=0
\end{equation}

\begin{equation}
\dot{\rho}_{_{dm}}+3H(1+w_{_{dm}}^{(e)})\rho_{_{dm}}=0
\end{equation}

By means of equations (15), (16), (21), and (22) one can obtain

\begin{equation}
w_{_{X}}^{(e)}=\gamma+w_{_{X}}~\left[
\frac{\frac{w_{_{X}}A_{s}}{w_{_{X}}-\gamma}+(1-A_{s})a^{3\eta(w_{_{X}}-\gamma)}
\left((1-\eta)\frac{\gamma}{w_{_{x}}} +\eta
\right)}{\left(\frac{\gamma(1-A_{s})}{w_{_{X}}A_{s}}~a^{3\eta(w_{_{X}}-\gamma)}
+\frac{w_{_{X}}}{w_{_{X}}-\gamma}
\right)\left((1-A_{s})~a^{3\eta(w_{_{X}}-\gamma)}+\frac{w_{_{X}}A_{s}}{w_{_{X}}-\gamma}
\right)} \right]
\end{equation}

\begin{equation}
w_{_{dm}}^{(e)}=\gamma+\left[
\frac{(\eta-1)w_{_{X}}A_{s}}{(1-A_{s})~a^{3\eta(w_{_{X}}-\gamma)}+\frac{w_{_{X}}A_{s}}{w_{_{X}}-\gamma}
} \right]
\end{equation}

Considering the spatially flat universe, the Friedmann equation
can be written as

\begin{equation}
3M_{p}^{2}H^{2}=\rho_{_{ch}}+\rho_{_{b}}
\end{equation}

where, $M_{p}$ is the reduced Plank mass and $\rho_{_{b}}$ is the
baryon matter density. The Friedmann equation can also be
expressed as

\begin{equation}
H(a)=H_{0}E(a)
\end{equation}

where

\begin{equation}
E(a)=\left[(1-\Omega_{b}^{0})a^{-3} \left\{
(1-A_{s})a^{-3(1+\alpha)\gamma}+\left(
\frac{w_{_{X}}A_{s}}{w_{_{X}}-\gamma}
\right)~a^{-3(1+\alpha)w_{_{X_{}}}} \right\}^{\frac{1}{1+\alpha}}
+\Omega_{b}^{0}a^{-3} \right]^{1/2}
\end{equation}

Then the density parameters of various components can be obtained,

\begin{equation}
\Omega_{X}=\Omega_{X}^{0}E^{-2}~a^{-3(1+\gamma)}\frac{\frac{\gamma}{w_{_{X}}}~(1-A_{s})+
\frac{w_{_{X}}A_{s}}{w_{_{X}}-\gamma}~a^{-3\eta(w_{_{X}}-\gamma)}}{\frac{\gamma}{w_{_{X}}}~(1-A_{s})+
\frac{w_{_{X}}A_{s}}{w_{_{X}}-\gamma} }
 \left(1+\frac{\gamma A_{s}}{w_{_{X}}-\gamma}
\right)^{1-\frac{1}{\eta}}\left[1-A_{s}
\left(1-\frac{w_{_{X}}}{w_{_{X}}-\gamma}~a^{-3\eta(w_{_{X}}-\gamma)}
\right) \right]^{\frac{1}{\eta}-1}
\end{equation}

\begin{equation}
\Omega_{dm}=(1-\Omega_{X}^{0}-\Omega_{b}^{0})E^{-2}~a^{-3(1+\gamma)}\left(1+\frac{\gamma
A_{s}}{w_{_{X}}-\gamma} \right)^{1-\frac{1}{\eta}}\left[1-A_{s}
\left(1-\frac{w_{_{X}}}{w_{_{X}}-\gamma}~a^{-3\eta(w_{_{X}}-\gamma)}
\right) \right]^{\frac{1}{\eta}-1}
\end{equation}

\begin{equation}
\Omega_{b}=\Omega_{b}^{0}E^{-2}a^{-3}
\end{equation}

\begin{figure}
\includegraphics[height=2.2in]{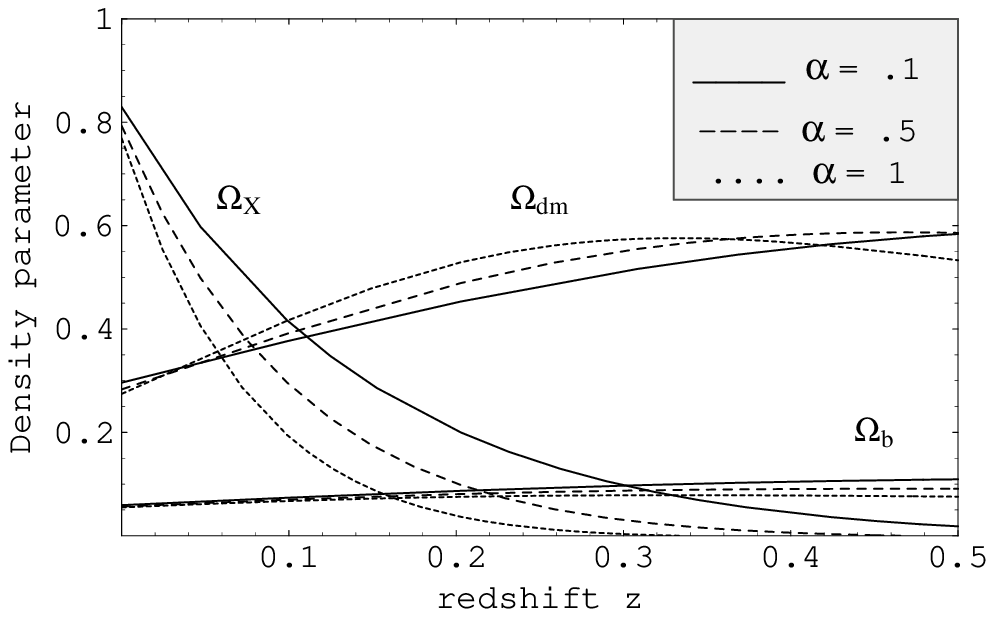}~~~~
\includegraphics[height=2.2in]{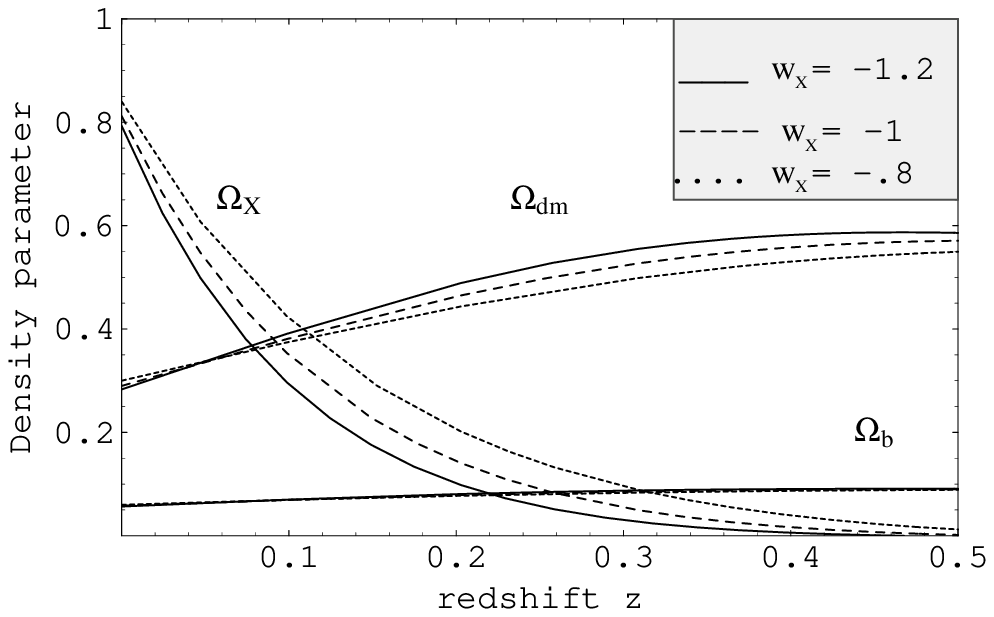}\\
\vspace{1mm} ~~~~~~~~~~~~Fig.1~~~~~~~~~~~~~~~~~~~~~~~~~~~~~~~~~~~~~~~~~~~~~~~~~~~~~~~~~~~~~~~~~~~~~~~~~~~~~~~~~~Fig.2\\

\vspace{7mm}

\includegraphics[height=2.2in]{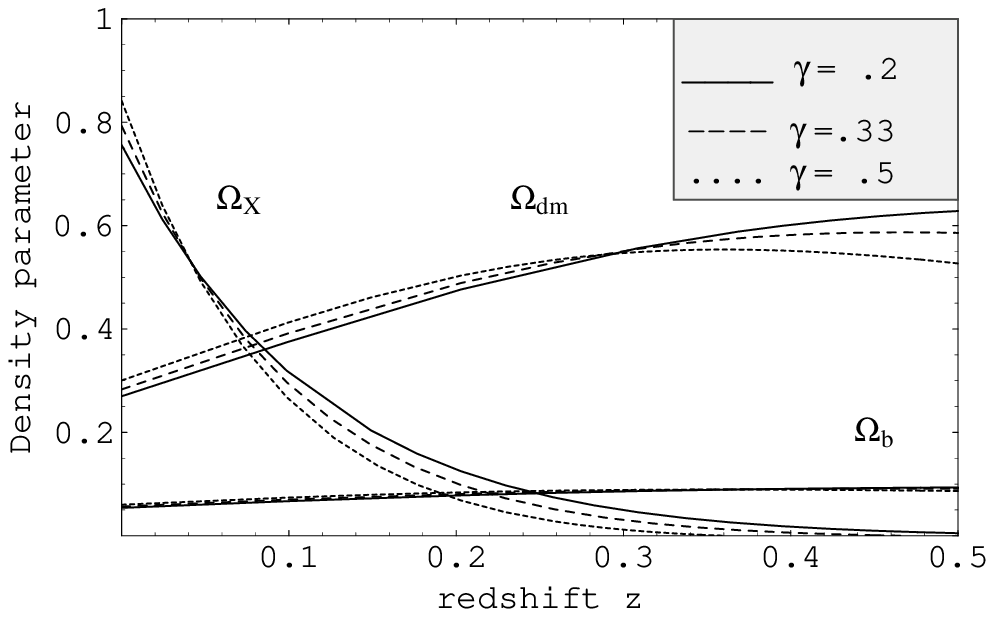}~~~~
\includegraphics[height=2.2in]{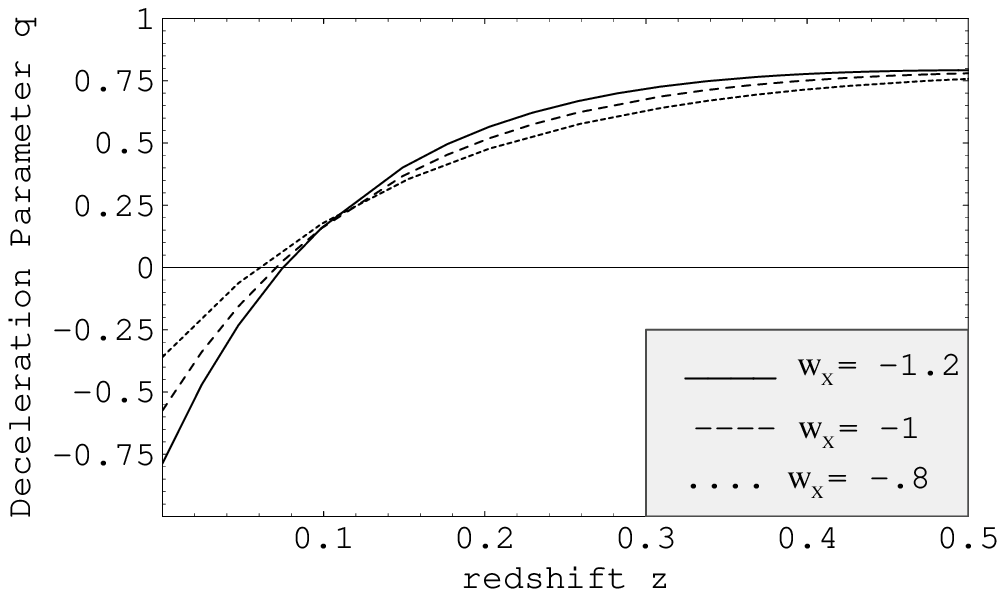}\\
\vspace{1mm} ~~~~~~~~~~~~~Fig.3~~~~~~~~~~~~~~~~~~~~~~~~~~~~~~~~~~~~~~~~~~~~~~~~~~~~~~~~~~~~~~~~~~~~~~~~~~~~~~~~~~~Fig.4\\

\vspace{7mm} {\bf Figs. 1 - 3} show the evolution of the density
parameters for various components $\Omega_{X},~\Omega_{dm}$ and
$\Omega_{b}$. The current density parameters used in the plots are
$\Omega_{ch}^{0}=0.25,~\Omega_{X}^{0}=0.7$ and
$\Omega_{b}^{0}=0.05$. In this case, Fig. 1 shows $w_{_{X}}$ and
$\gamma$ fixed and $\alpha$ is varied; Fig. 2
shows\\\hspace{-7.7cm} $\alpha$ and $\gamma$ fixed and $w_{_{X}}$
is varied; Fig. 3 shows $\alpha$ and $w_{_{X}}$ fixed and $\gamma$ is varied. \\
\hspace{-.6cm} {\bf Fig. 4} shows the evolution of the
deceleration parameter $q(z)$. The current density parameters
used in the plots are
$\Omega_{ch}^{0}=0.25,$\\\hspace{-7.3cm}~$\Omega_{X}^{0}=0.7$ and
$\Omega_{b}^{0}=0.05$. In this case, $w_{_{X}}$ are taken to be
$-0.8,~ -1$ and $-1.2$~.\\

\vspace{6mm}

\end{figure}

The density evolution of the NMCG model is given in Figures 1, 2
and 3. The current density parameters used in the plots are
$\Omega_{dm}^{0}$ =0.25, $\Omega_{X}^{0}$ =0.7 and
$\Omega_{b}^{0}$ =0.05. In figure 1, we show the cases having the
common equation-of-state parameters $w_{_{X}} =-1.2$ and
$\gamma=1/3$, while the parameter $\alpha$ are taken to be 0.1,
0.5 and 1 respectively. In figure 2, the evolution of the density
parameter is studied keeping the parameters $\alpha$ fixed at 0.5
and $\gamma=1/3$ and the equation-of-state parameters $w_{_{X}}$
are taken to be $-0.8,~ -1$ and $-1.2$. Evolution of density with
varying $\gamma$ is depicted in figure 3. In this figure, the
parameter $\alpha$ is fixed at 0.5, equation-of-state parameter
$w_{_{X}}$ is taken as $-1.2$, and values of $\gamma$ are taken
to be 1/5, 1/3, and 1/2.  The acceleration of the Universe is
evaluated by the deceleration parameter

\begin{equation}
q=-\frac{\ddot{a}}{aH^{2}}
\end{equation}

In the NMCG model, the deceleration parameter comes out to be

\begin{equation}
q=\frac{1}{2}~(1+3w_{_{X}}\Omega_{X}+3\gamma\Omega_{dm})
\end{equation}

Evolution of the deceleration parameter $q$ is shown in figure 4.

\section{\normalsize\bf{Concluding Remarks}}

The new modified Chaplygin gas model is regarded as a unification 
of dark energy and dark matter (with general form of matter i.e.,
$\gamma\neq 0$). This model interpolates between radiation at
early stage and $\Lambda$CDM at late stage. The unification of
dark energy and dark matter should accommodate the
quintessence-like ($-1<w_{_{X}}<-1/3$) and phantom-like
($w_{_{X}}<-1$) dark energy. From figures 1 - 3, we have seen that
$\Omega_{dm}$ first increases and then decreases to a constant
value and $\Omega_{X}$ decreases to a constant value but in every
stage in the evolution of the universe, the sum remains
approximately equal to 1. From the figure 4,  it can be seen
that  the deceleration parameter $q$ decreases from positive value
to negative value i.e., the evolution of the universe demands
early deceleration and late acceleration. From equation (32) it
seems that the positive part is larger in NMCG than NGCG
($\gamma=0$). Therefore, NMCG is able to describe the deceleration
part of the universe in a larger range than NGCG.\\

{\bf Acknowledgement:}\\\\
One of the authors (UD) is thankful to BESU, India for
providing a research project grant (No. DRO-2/6858).\\

{\bf References:}\\
\\
$[1]$ S. J. Perlmutter et al, {\it  Astrophys. J.} {\bf 517} 565
(1999); A. G. Rieses et al, {\it Astron. J.} {\bf 116} 1009
(1998); P. M. Garnavich et al, {\it Astrophys. J.} {\bf 509} 74
(1998); G. Efstathiok et al, {\it astro-ph}/9812226. \\
$[2]$ B. Ratra and P. J. E. Peebles, {\it Phys. Rev. D} {\bf 37}
3406 (1988); R. R. Caldwell, R. Dave and P. J. Steinhardt, {\it
Phys. Rev. Lett.} {\bf 80} 1582 (1998).\\
$[3]$ S. Bridle, O. Lahav, J. P. Ostriker and P. J. Steinhardt,
{\it Science} {\bf 299} 1532 (2003); C. Bennet et al, {\it
astro-ph}/0302207; D. N. Spergel et al, {\it astro-ph}/0302209; V.
Sahni and A. A. Starobinsky, {\it Int. J. Mod. Phys. D}
{\bf 9} 373 (2003); T. Padmanabhan, {\it hep-th}/0212290.\\
$[4]$ V. Sahni and A. A. Starobinsky, {\it Int. J. Mod. Phys. D}
{\bf 9} 373 (2003).\\
$[5]$ Armendariz-Picon, T. Damour and V. Mukhanov, {\it Phys.
Lett. B} {\bf 458} 209 (1999) ; C. Armendariz-Picon, V. Mukhanov
and P. J. Steinhardt, {\it Phys. Rev. D} {\bf 63} 103510 (2001). \\
$[6]$ T. Chiba, {\it Phys. Rev. D} {\bf 66} 063514 (2002). \\
$[7]$ M. C. Bento, O. Bertolami and A. A. Sen, {\it Phys. Rev. D}
{\bf 67} 063511 (2003).\\
$[8]$ R. R. Caldwell, {\it Phys. Lett. B} {\bf 545} 23 (2002). \\
$[9]$ L. A. Boyle, R. R. Caldwell and M. Kamionkowski, {\it Phys.
Lett. B} {\bf 545} 17 (2002). \\
$[10]$ A. Melchiorri, L. Mersini, C. J. Odman and M. Trodden,
{\it Phys. Rev. D} {\bf 68} 043809 (2003).\\
$[11]$ A. Kamenshchik, U. Moschella and V. Pasquier, {\it Phys.
Lett. B} {\bf 511} 265 (2001).\\
$[12]$ V. Gorini, A. Kamenshchik and U. Moschella, {\it Phys. Rev.
D} {\bf 67} 063509 (2003); U. Alam, V. Sahni , T. D. Saini and
A.A. Starobinsky, {\it Mon. Not. Roy. Astron. Soc.} {\bf 344}, 1057 (2003) .\\
$[13]$ M. C. Bento, O. Bertolami and A. A. Sen, {\it Phys. Rev. D}
{66} 043507 (2002) .\\
$[14]$ H. B. Benaoum, {\it hep-th}/0205140. \\
$[15]$ U. Debnath, A. Banerjee and S. Chakraborty, {\it Class.
Quantum Grav.} {\bf 21} 5609 (2004). \\
$[16]$ Z. K. Guo and Y. Z. Zhang, {\it Phys. Lett. B} {\bf 645} 326 (2007); {\it astro-ph}/0506091.\\
$[17]$ M. C. Bento, O. Bertolami and A. A. Sen, {\it Phys. Lett.
B} {\bf 575} 172 (2003). \\
$[18]$ Z. K. Guo and Y. Z. Zhang, {\it astro-ph}/0509790; G.
Sethi, S. K. Singh, P. Kumar, D. Jain and A. Dev, {\it Int. J.
Mod. Phys. D} {\bf 15} 1089 (2006); {\it astro-ph}/0508491; X. -Y.
Yang, Y. -B. Wu, J. -B. Lu and S. Li, {\it Chin. Phys. Lett.} {\bf 24} 302 (2007);
U. Debnath, {\it Astrphys. Space Sci.} {\bf 312} 295 (2007).\\
$[19]$ X. Zhang, F. -Q. Wu and J. Zhang, {\it JCAP} {\bf
01} 003 (2006).\\
$[20]$ V. Sahni and A. Starobinsky, {\it Int. J. Mod. Phys. D}
{\bf 9} 373 (2000); V. Sahni, {\it Class. Quantum Grav.} {\bf 19}
3435 (2002).\\
$[21]$ M. C. Bento, O. Bertolami and A. A. Sen, {\it Phys. Rev.
D} {\bf 70} 083519 (2004); W. Zimdahl and J. C. Fabris, {\it
Class. Quantum Grav.} {\bf 22} 4311 (2005).\\

\end{document}